# Stress deformations and structural quenching in charge-ordered $Sm_{0.5}Ca_{0.5}MnO_3$ thin films

E. Rauwel, W. Prellier and B. Mercey

Laboratoire CRISMAT, CNRS UMR 6508, 6, Bd du Maréchal Juin, F-14050 Caen cedex, FRANCE

S. de Brion and G. Chouteau

Grenoble High Magnetic Field Laboratory, CNRS and MPI-FKF, BP 166, F-38042 Grenoble cedex 9, FRANCE

Corresponding author: rauwelerwan@yahoo.fr

## Abstract

$Sm_{0.5}Ca_{0.5}MnO_3$ manganite thin films with charge ordering (CO) and colossal magnetoresistance were synthesized by pulsed laser deposition on (100)-$SrTiO_3$ and (100)-$LaAlO_3$ substrates. We first compare the structural modifications as functions of the substrate and film thickness. Secondly, transport properties in magnetic fields up to 24T were studied and the temperature-field phase diagram describing the stability of the CO state was established. This enables us to compare the thin film to the bulk material. We show that some substrate induced structural modifications exist as a result of which the CO melting magnetic field is greatly reduced. Moreover, no modification of the lattice parameters is observed with temperature decrease. We then propose an explanation based on the quenching of the thin film's unit cell which then adopts the *in-plane* lattice parameters of the substrate and thus prevents the complete setting in of the CO state at low temperature.

# I. INTRODUCTION

Over the past few years, perovskite type manganites such as $R_xA_{1-x}MnO_3$ (R = rare earth elements and A = alkaline earth ions) have been extensively investigated due to their colossal magnetoresistance properties (CMR)[1,2,3] which exist around the ferromagnetic transition (x~ 0.3) or due the melting of the charge ordered state (x~ 0.5). It has been shown in bulk compounds that these CMR properties are affected by perturbations like chemical pressure[4,5] or external pressure[6]. For instance, Hwang et al.[4] showed that there exists a direct relationship between the average ionic radius $<r_a>$ and the Curie temperature $T_C$ in $La_{0.7}A_{0.3}MnO_3$. Fractionally doped, $R_{0.5}Ca_{0.5}MnO_3$ compounds present charge ordering (CO) in $Mn^{3+}$, $Mn^{4+}$ sublattices[7]. This charge ordering takes place in two different Mn sublattices (i.e. a long-range ordering of $Mn^{3+}$ and $Mn^{4+}$ ions) below the CO temperature. The CO state is highly insulating, but can be melted (by inducing a ferromagnetic metallic state)[8] by applying a magnetic field[9]. It has to be pointed out that this CO transition is associated with a large lattice deformation[9].

The work presented here is part of a more general study which aims to understand the effects of both the internal pressure and the substrate pressure on the CO state when the compound is synthesized in the thin film form. The internal pressure is governed by the size of the rare earth element, whereas the substrate stress depends on its lattice parameters and the film's thickness. We have already shown that for $Pr_{0.5}Ca_{0.5}MnO_3$[12,13] and $Nd_{0.5}Ca_{0.5}MnO_3$[14,15] thin films, a lower CO melting magnetic field ($H_C$) was obtained. This critical magnetic field was 6 times lower than that of the bulk compound. We have attributed this effect to the substrate-induced strain and have established an analogy between the strain and the stability of the CO state. Internal pressure has a strong influence on these half doped magnanites. In the case of bulk materials, $H_C$ increases with the decrease of the average ionic radius[16]. In particular, when going from $Pr_{0.5}Ca_{0.5}MnO_3$ $<r_a>$=1.1795Å[17] to $Sm_{0.5}Ca_{0.5}MnO_3$ $<r_a>$=1.156Å[17] there is an important increase of the melting magnetic field and a field of 25T is necessary to melt the CO in $Pr_{0.5}Ca_{0.5}MnO_3$ whereas 60T is required[18] for $Sm_{0.5}Ca_{0.5}MnO_3$ at a temperature of 50K. These values of the magnetic fields correspond to the maximum ones that should be applied in order to melt the CO state in such bulk compounds.

In the present work we concentrate on $Sm_{0.5}Ca_{0.5}MnO_3$ thin films deposited on two different substrates: $SrTiO_3$ and $LaAlO_3$. In order to elucidate the phenomena of the substrate-induced strain, we have performed low temperature characterisation measurements on all the films (from the structural as well as electrical points of view). This allows us to understand how strain influences the properties of the film, in particular the stability of the CO state. We have chosen $Sm_{0.5}Ca_{0.5}MnO_3$ because we expect it to have more pronounced effects such as the decrease of the critical magnetic field as compared to $Pr_{0.5}Ca_{0.5}MnO_3$ and $Nd_{0.5}Ca_{0.5}MnO_3$ since Sm has a smaller radius. Moreover the deformations of the unit cell should be higher in $Sm_{0.5}Ca_{0.5}MnO_3$ than in $Pr_{0.5}Ca_{0.5}MnO_3$ and $Nd_{0.5}Ca_{0.5}MnO_3$.

## II. EXPERIMENTAL

A dense target of $Sm_{0.5}Ca_{0.5}MnO_3$ was sintered by the standard ceramic method. Thin films of $Sm_{0.5}Ca_{0.5}MnO_3$ were grown in situ using pulsed laser deposition on cubic (100)-$SrTiO_3$ (a = 3.905Å) and pseudocubic (100)-$LaAlO_3$ (a = 3.789Å) substrates. The films were deposited using a KrF laser (?=248mm)[12-15]. The substrates were maintained at a constant temperature ($T_S$) ranging from 620°C to 750°C during the deposition. This was carried out in a pure oxygen pressure of around 400mTorr. The deposition rate was 3Hz and the energy density was close to 2J/cm². The samples were then slowly cooled down to room temperature with an oxygen pressure of 500mbar. These conditions ensure that optimal oxidation of the film takes place. This is confirmed by the fact that no change was observed in the lattice parameters of the films after an oxygen based annealing at 700°C for 10 hours. The overall chemical composition of the film was checked by energy dispersive spectroscopy (EDS) and it corresponds to the nominal one in the limit of accuracy ($Sm_{0.5\pm0.02}Ca_{0.5\pm0.02}MnO_x$). Structural characterizations were performed at room temperature using a Seifert XRD 3000P diffractometer ( Cu $K_a$, ? = 1.54056Å) and a JEOL 2011 FX electron microscope (2.4Å resolution) coupled with EDS analysis. A X'Pert Phillips diffractometer was used for the low temperature (100K) structural characterizations. Electrical resistivity measurements ($\rho$) of the films were performed as a function of temperature (4K-300K) in magnetic fields (up to 24T) at the Grenoble High Magnetic Field Laboratory using four-probe contacts. The magnetic field was varied at a constant rate of 0.025T/sec in the ramping up and down modes.

III. RESULTS AND DISCUSSION

a. Structural study

The best crystallization was obtained with the substrate temperature $T_S$ ranging from 650°C to 670°C and an oxygen pressure of 400mTorr. Figure 1a and 1b show a typical x-ray diffraction (XRD) $\theta$-$2\theta$ scan recorded for films grown on SrTiO$_3$ and on LaAlO$_3$ respectively. The sharp and intense peaks observed in the diffractograms at room temperature indicate that the films are highly crystallized. This is also confirmed by the low value (0.1°) of the full-width at half maximum of the rocking curve recorded around the 101 reflection (see inset of Fig.1a). The main diffraction peaks are located at $2\theta = 47.76°$ for SrTiO$_3$ (Fig 1a) and $2\theta = 47.54°$ for LaAlO$_3$ (Fig 1b). They correspond to an *out-of-plane* lattice parameter, which is a multiple of 1.902Å for SrTiO$_3$ and 1.911Å for LaAlO$_3$. Sm$_{0.5}$Ca$_{0.5}$MnO$_3$ is orthorhombic (Pnma space group)[19] with $a = 5.4159$Å $\approx a_C\sqrt{2}$, $b = 7.5485$Å $\approx 2a_C$, $c = 5.3601$Å $\approx a_C\sqrt{2}$ (where $a_C$ refers to the ideal cubic perovskite = 3.900Å) and the $d_{202}$ and $d_{040}$ distance of the bulk is equal to 1.904Å and 1.887Å respectively. Therefore the diffraction peak located around $2\theta = 47°$ can be indexed either as a 202 or a 040 reflection. The epitaxial orientation of the films were investigated by $\Phi$–scans for an asymmetrical reflection. An example of such a pole figure is given in figure 2 for a film grown on SrTiO$_3$. Four peaks are clearly observed at 90° from each other, indicating a four-fold symmetry as expected for the orthorhombic structure of Sm$_{0.5}$Ca$_{0.5}$MnO$_3$[19]. This well defined pattern is an evidence of the *in-plane* texture of the Sm$_{0.5}$Ca$_{0.5}$MnO$_3$ phase. Similar scans on other films confirm that the Sm$_{0.5}$Ca$_{0.5}$MnO$_3$ layer grows epitaxially on both substrates used in this study.

An investigation using electron diffraction shows that $Sm_{0.5}Ca_{0.5}MnO_3$ is mainly [**101**]-oriented on both substrates (fig. 3 and fig. 4) with the presence of large 90° [**101**]-oriented domains in the substrate plane. We nevertheless notice the presence of a few small [**010**]-oriented domains in the film. $Pr_{0.5}Ca_{0.5}MnO_3$ and $Nd_{0.5}Ca_{0.5}MnO_3$ thin films deposited on $SrTiO_3$ are mostly [**010**]-oriented whereas they are mainly [**101**]-oriented when deposited on $LaAlO_3$ substrate[12,15]. This difference in strain results from the minimum lattice mismatch which exists between the film and the substrate for these orientations[13]. In the SMCO films, we have calculated the lattice mismatch using the usual formula: $\mathbf{s} = 100*(a_S-a_F)/a_S$, where $a_S$ and $a_F$ refer to the lattice parameter of the substrate and the film respectively. The values are reported in Table I for the [**010**]-and [**101**] orientations on $LaAlO_3$ and $SrTiO_3$. Despite a lower mismatch at room temperature between $SrTiO_3$ parameter and the 101 $Sm_{0.5}Ca_{0.5}MnO_3$ parameter (+2.40%), the film actually grows mainly in the [**101**] direction on the $SrTiO_3$ substrate. In this case, the mismatch between 001 $SrTiO_3$ parameter and the 010 $Sm_{0.5}Ca_{0.5}MnO_3$ parameter is positive (+3.38%); this induces a tensile strain along [010] direction and a compressive strain along the [101] direction. Thus the *b* parameter of the orthorhombic cell increases and the *a* and *c* parameters decrease.

For films grown on $LaAlO_3$, the mismatch at room temperature between the parameter of the substrate and the 101 $Sm_{0.5}Ca_{0.5}MnO_3$ parameter at room temperature is -0.60% while it is +0.42% for the 010 $Sm_{0.5}Ca_{0.5}MnO_3$ parameter. TEM study shows the presence of very large 90° [**101**]-oriented domains in the substrate's plane. The film has chosen an orientation with a lower lattice mismatch and an almost perfect match exists between the $c/\sqrt{2}$ $Sm_{0.5}Ca_{0.5}MnO_3$ parameter (3.791Å) and the $LaAlO_3$ parameter (3.789Å). The mismatch is now positive but much lower as compared to the $SrTiO_3$ substrate. The cell is less deformed in this case. As for films grown on $SrTiO_3$, there are a few small [**010**]-oriented domains.

To understand better the effect of the substrate on the films' structural properties, we have studied the influence of the film thickness on the cell parameters. The *in-plane* parameter was measured with a Philips X'Pert X-ray diffractometer using the $(103)_C$ reflection. For $Sm_{0.5}Ca_{0.5}MnO_3$ films grown on $SrTiO_3$ with $T_S$ ranging from 620°C to 670°C the *out-of-plane* parameter is constant and a pseudocubic symmetry persists. For a film thickness of 1110Å grown with $T_S = 650°C$, the *out-* *of-plane* parameter $a_\perp$ is 3.758Å and in-plane parameter $a_{//}$ is 3.843Å (See Table II). As the thickness of the film increases, the *out-of-plane* lattice parameter $a_\perp = d_{101}$ increases and at the same time, the *in-plane* parameter $a_{//} = d_{010}$ decreases. Thus for 2000Å $Sm_{0.5}Ca_{0.5}MnO_3$ thin films grown on $SrTiO_3$ with $T_S$ = 650°C we measured $a_\perp$ =3.816Å and $a_{//}$ = 3.806Å. The tendency seems to be the same for $Pr_{0.5}Ca_{0.5}MnO_3$[13] and $Nd_{0.5}Ca_{0.5}MnO_3$[15] on $SrTiO_3$ substrates. We also observe a decrease of the "orthorhombic" character of the structure (the difference in the *in-plane* and the *out-of-plane* parameters becomes smaller) as the thickness of the film increases.

For $Sm_{0.5}Ca_{0.5}MnO_3$ films grown on $LaAlO_3$, the *out-of-plane* parameter remains almost constant with the change in thickness, suggesting that strains are nearly relaxed even for a small thickness. This is easily explained by the low mismatch. We were unable to obtain the *in-plane* lattice parameter for these films due to the twinning of the $LaAlO_3$ substrate.

b. Physical properties

We now analyse the electrical transport measurements of the best crystallised films. All the films show a semiconducting behaviour in the temperature range 4-300*K* with an anomaly around 276K corresponding to $T_{CO}$ (CO transition). Note that this transition is very close to the bulk value[20] (Fig. 5). In fact, this anomaly in the *?(T)* curves is cleared (inset of figure 5) when the resistance is plotted on the logarithmic scale versus the inverse of the temperature. It is well known that in bulk SMO magnetic field as high as 60T is necessary to melt the CO state and promote the transition towards the ferromagnetic metallic state[11]. For our thin films, we have used a resistive magnet up to 23*T* at the Grenoble High Magnetic Field Laboratory. Electrical resistivity as a function of magnetic field recorded at different temperatures is presented in figure 6 for 2000Å thick films grown on $SrTiO_3$ (Fig. 6a) and $LaAlO_3$ (Fig. 6b). In both films, the resistivity on a logarithmic scale presents a decrease at the critical field $H_C$. This decrease is particularly huge below 100K: the CMR effect calculated as a ratio of *?(H=0)/?(H=13T)*, is more than $10^4$. This field-induced insulator to metal transition, which occurs due to the collapsing of the CO state, takes place below $T_{CO}$. There is a strong hysteresis between the increasing field ($H_C^+$) and decreasing field ($H_C^-$) behaviour as observed in bulk materials[8,9,21] but the field values are very different. In figure 7 the phase diagram is plotted in the temperature-magnetic field plane, and is compared to the bulk material[18].

Firstly, in the thin films, the critical magnetic field is always lower (at 50K, $H_C$ = 60T for the bulk compound[18] whereas $H_C$ = 11T and $H_C$ = 16T for SrTiO$_3$ and LaAlO$_3$ substrates respectively). Note also that the hysteretic region is smaller than in the bulk material and is more pronounced at low temperature. This large hysteretic region is characteristic of a first order transition that has been extensively studied in Nd$_{0.5}$Sr$_{0.5}$MnO$_3$ compounds[9]. Secondly, the shape of the *H-T* phase diagram is totally different as compared to the bulk single crystal. The critical magnetic field is found to be minimum at 25K (on both substrates), whereas for the Pr$_{0.5}$Ca$_{0.5}$MnO$_3$ and Nd$_{0.5}$Ca$_{0.5}$MnO$_3$ films, the minimum values are found respectively to be 50K and 75K. However the critical magnetic field is lower for the film deposited on SrTiO$_3$ than for the film on LaAlO$_3$. Previously we have shown that the substrate-induced strains are higher for SrTiO$_3$ substrates, with a mismatch of +3.38% than for LaAlO$_3$ with a mismatch of +0.42%. Hence the structural deformation is larger in the case of the SrTiO$_3$ substrate. This means that such a structural deformation decreases the stability of the CO state, which in turn reduces the critical field for the CO melting.

c. Charge order stability and structural change

It is important to keep in mind that the CO transition is associated with large structural changes in the bulk[8,9]. In order to verify the existence of this structural transition the thin films, we have performed low temperature x-ray diffraction below $T_{CO}$ (100K i.e. in the CO state) on the two films: one grown on SrTiO$_3$ and the other grown on LaAlO$_3$. The resulting data is presented in figure 8. On both substrates, the *out-of-plane* lattice parameter of the film remains almost at the same value after cooling. For example, for the film on LaAlO$_3$, the *out-of-plane* is equal to 3.824Å at 298K and 3.821Å at 100*K*. Taking into account that the substrate-induced strains are stronger in the plane of the substrate and that the *out-of-plane* parameter is constant for all temperatures between 100K and 298K, we can then suppose that the *in-plane* parameter doesn't vary either. Keeping in mind that CO transition occurs around 280K, this experiment demonstrates the quenching of the film's structure under cooling. In fact whatever be the temperature, the lattice parameters of the $Sm_{0.5}Ca_{0.5}MnO_3$ do not change from their room temperature values (Fig. 8). This prevents the establishing of a complete structural transition associated with the CO state. The CO state is consequently less stable and it is therefore easier to induce a metal-to-insulator transition towards the ferromagnetic state.

If we compare these results to those obtained from $Pr_{0.5}Ca_{0.5}MnO_3$ and $Nd_{0.5}Ca_{0.5}MnO_3$ thin films[12,14,15,22], in both cases, the structure of the film was subjected to a compressive strain along the [010] direction and a tensile strain along the [101] direction which induce a decrease of the $b$ parameter and an increase of the $a$ and $c$ parameters. This distortion induces the same structural deformation which takes place at the structural transition temperature ($T_{CO}$) that stabilizes the CO state. This explains why a lower thickness increases the critical magnetic field ($H_C$)[15,13].

In the case of $Sm_{0.5}Ca_{0.5}MnO_3$, the films are [**101**]-oriented with a tensile strain in the substrate plane ($SrTiO_3$ and $LaAlO_3$). So the structural deformation is different in this case. In contrast with the $Pr_{0.5}Ca_{0.5}MnO_3$ and $Nd_{0.5}Ca_{0.5}MnO_3$[23] compounds, the substrate-induced strains provoke an increase of the $b$ parameter and a decrease of the $a$ and $c$ parameters. This distortion creates a structural deformation which is opposite to the structural transition which takes place at $T_{CO}$. The comparison between the thin film parameters and the bulk parameters[24] as a function of temperature (Fig. 9) enables a better understanding of the behaviour of the films. The *out-of-plane* parameters are noted by a diamond (?) and the *in-plane* parameter by a circle (?). The lines indicate that the film's parameters do not vary with temperature, as shown previously.

In the case of the $SrTiO_3$ substrate, the deformation is more important because of the higher mismatch. For this reason, we observe a huge decrease of $H_C$ in the $Sm_{0.5}Ca_{0.5}MnO_3$ thin film grown on $SrTiO_3$ and 11T is enough to melt the CO state (Fig. 7), as compared to 60T required for the bulk material[18]. For films grown on $LaAlO_3$, the decrease of $H_C$ is less important and $H_C = 16T$. These results point out that the decrease of $H_C$ is closely linked to the distortion induced by the substrate, as well as the structural quenching, induced by the substrate, when the temperature is lowered. In fact, if the structural transition doesn't take place, the stability of the charge order is decreased and the corresponding melting magnetic field is decreased.

IV. CONCLUSION

In summary, we have grown $Sm_{0.5}Ca_{0.5}MnO_3$ films on two different substrates ($SrTiO_3$ and $LaAlO_3$) and have studied the effects of the strains on the structural and transport properties. The magnetic field required to melt the insulating charge order (CO) state and to transform it into the ferromagnetic metallic state is greatly reduced compared to the bulk material.

We have shown that this is a consequence of two phenomena: firstly, the lattice mismatch between the [**101**]-orientated $Sm_{0.5}Ca_{0.5}MnO_3$ film and the substrate produces a deformation which is opposite to the one induced by the charge ordered structural transition at $T_{CO}$. This distortion reduces the stability of the charge order in the compound. Secondly, the substrate induces a structural quenching of the CO state: it freezes the unit cell into the room temperature configuration and prevents the complete setting in of the CO state. We thus conclude that the decrease of the critical magnetic field ($H_C$) is essentially due to a structural effect.

The authors thank Dr. D. Pelloquin and Miss. M. Lozier for low temperature X-ray diffraction measurements, Dr. C. Dubourdieu and P. Singh-Rauwel for fruitful discussion and careful reading of this article. The GHMFL is 'Laboratoire conventionné à l'université Joseph Fourier'. Partial support of Indo-French Center for the Promotion of Advanced Research (CE-FIPRA/IFCPAR) under Project N°2808-1 is greatly acknowledged.

Table I. Lattice mismatches between the substrate lattice parameters and the lattice parameters of Sm$_{0.5}$Ca$_{0.5}$MnO$_3$ bulk compound at room temperature.

| | Lattice mismatch with SrTiO3 substrate a = 3,905Å | Lattice mismatch with LaAlO3 substrate a = 3,789Å |
|---|---|---|
| 101 Sm0,5Ca0,5MnO3 parameter a = 3,811Å | **2,40%** | -0,60% |
| 020 Sm0,5Ca0,5MnO3 parameter a = 3,773Å | 3,38% | **0,42%** |

Table II. Evolution of the lattice parameters with the film thickness for $Sm_{0.5}Ca_{0.5}MnO_3$ films on SrTiO$_3$ and LaAlO$_3$

| Thickness of the film grown at 650°C | 2000Å | 1110Å |
|---|---|---|
| out-of-plane parameter ($a_\perp$) | **$a_\perp$ = 3,816Å** | $a_\perp$ = 3,758Å |
| In-plane parameter ($a_{//}$) | $a_{//}$ = 3,806Å | **$a_{//}$ = 3,843Å** |

.

RAUWEL et al.

Figures Captions:

Figure 1: Room temperature T-2T XRD pattern of typical $Sm_{0.5}Ca_{0.5}MnO_3$ films grown at 650°C (a): on $SrTiO_3$ (b): on $LaAlO_3$. Inset of Fig. 1a : rocking curve (?-scan) of the (002) reflection of the film. Note the sharpness and the high intensity of the peaks.

Figure 2: Pole figure of the $Sm_{0.5}Ca_{0.5}MnO_3$ grown at 650°C on $SrTiO_3$ recorded around the 100 or 001.

Figure 3: (a) ED of a plane view for a $Sm_{0.5}Ca_{0.5}MnO_3$ on $SrTiO_3$ showing [101] axis perpendicular to the substrate plane. (b) Corresponding plane view HREM image showing the two orientations in the film.

Figure 4: (a) ED of a plane view for a $Sm_{0.5}Ca_{0.5}MnO_3$ on $LaAlO_3$ showing two domains with the [020] axis in the substrate plane. (b) Corresponding plane view HREM image with the [020] axis perpendicular to the substrate plane.

Figure 5: *r(T)* under 0T (black line) and 7T (doted line), we note the anomaly associate with $T_{CO}$, marked by an arrow. The inset depicts the evolution of the logarithmic resistance versus the inverse of the temperature under 0T (black line) and 7 T (doted line). Note the anomaly associated with $T_{CO}$ marked by an arrow).

Figure 6: Magnetoresistance at different temperatures for 2000Å films grown at 650°C on $SrTiO_3$ (a) and $LaAlO_3$ (b). Runs in increasing field and decreasing field are indicated by arrows.

Figure 7: Phase diagram for $Sm_{0.5}Ca_{0.5}MnO_3$ films on $SrTiO_3$ (a) and $LaAlO_3$ (b). $H_C^+$ and $H_C^-$ are taken at the inflection points in the *?(H)* curves of figure 6 for the up and down sweeps respectively.

Figure 8: T-2T XRD patterns, recorded at 100K and 298K, for typical $Sm_{0.5}Ca_{0.5}MnO_3$ films deposited at 650°C on $SrTiO_3$ (a) and $LaAlO_3$ (b) substrates.

Figure 9: Evolution of the lattice parameters of the single crystal as a function of the temperature (From Ref. 23 Tomioka *et al.*). The frozen parameters of thin films grown on $SrTiO_3$ (dashed line and open symbol) and $LaAlO_3$ (doted line and full symbol) are indicated: (?) symbolize respectively the *out-of-plane* parameters of films, (?) symbolizes *in-plane* parameters of film. Lines which represent the films' parameters that remain constant for all temperatures are only a guide for the readers.

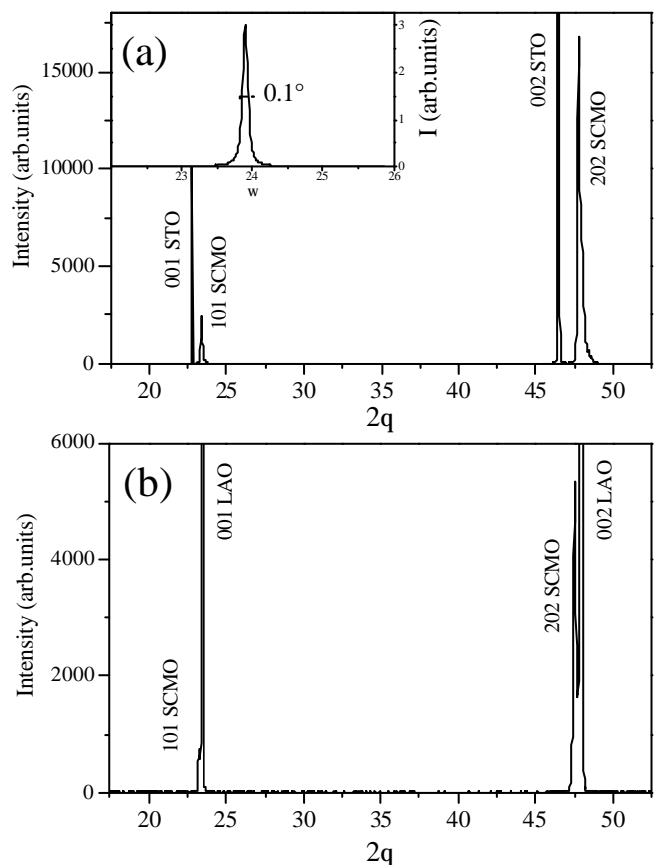

Figure 1

RAUWEL et al.

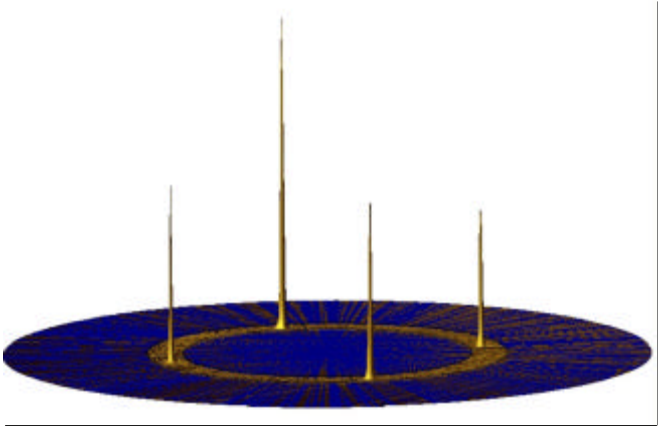

Figure 2

RAUWEL et al.

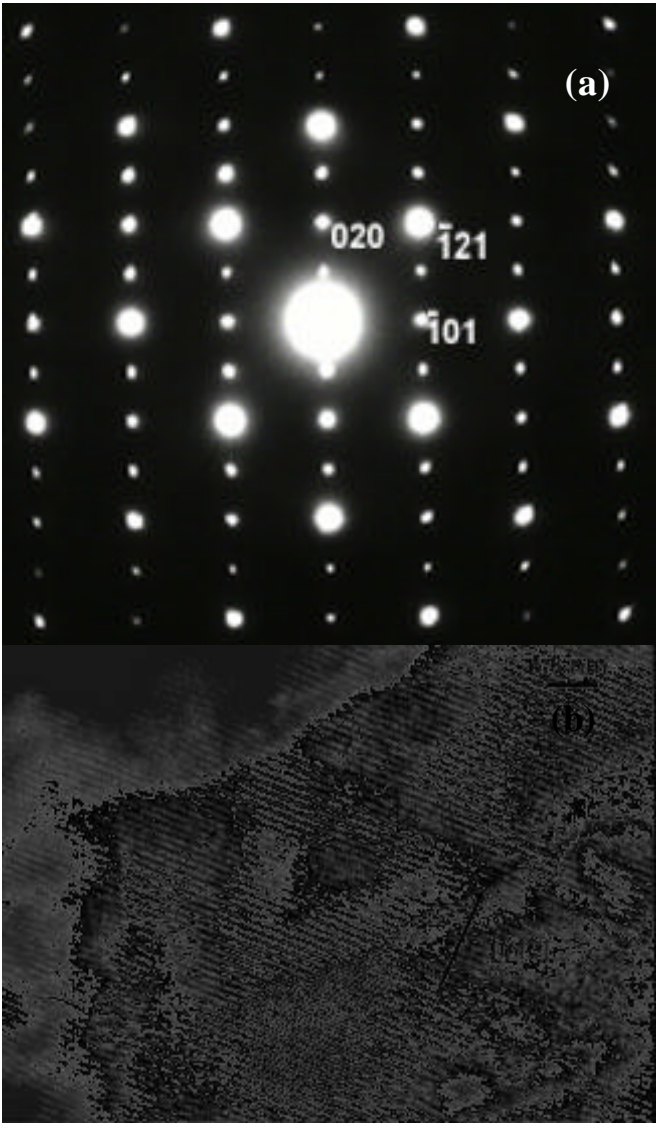

Figure 3

RAUWEL et al.

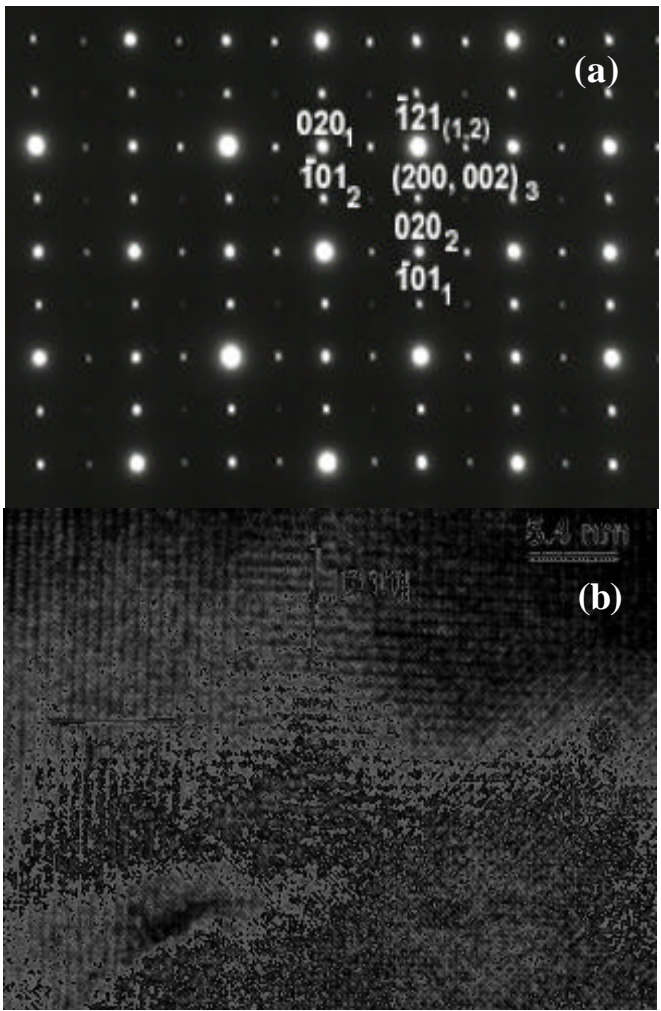

Figure 4

RAUWEL et al.

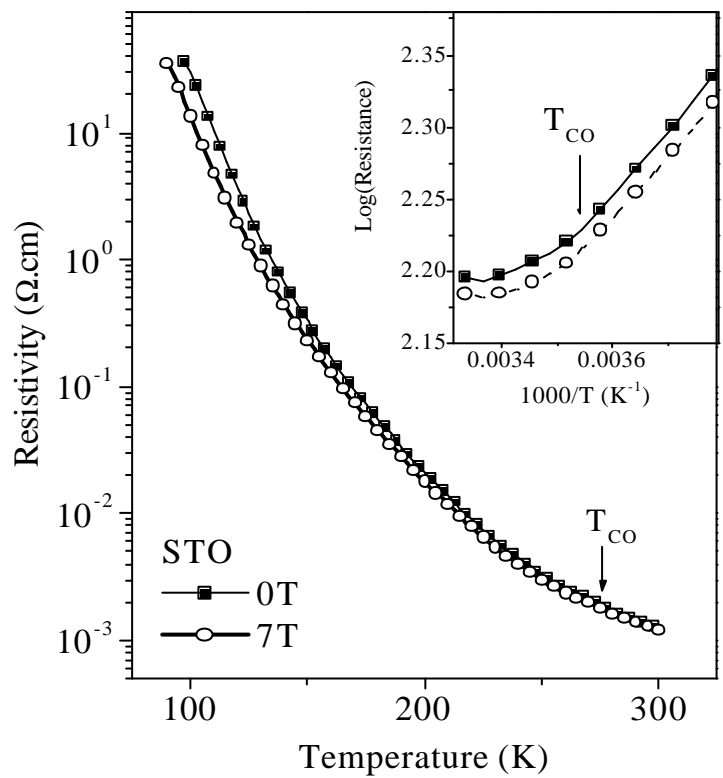

Figure 5

RAUWEL et al.

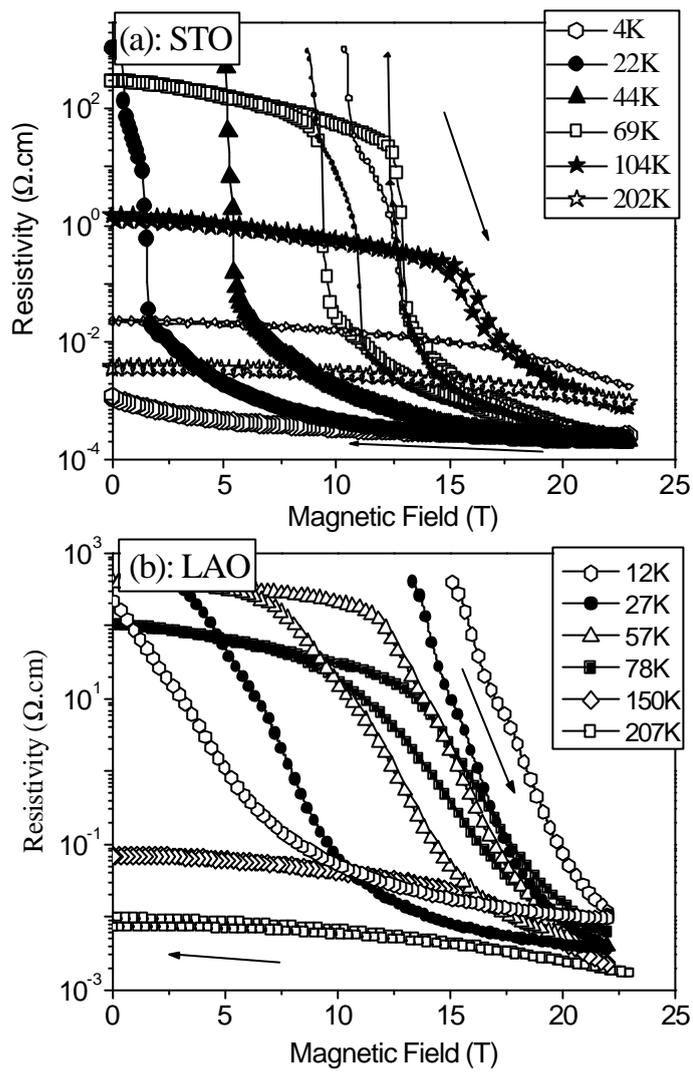

Figure 6

RAUWEL et al.

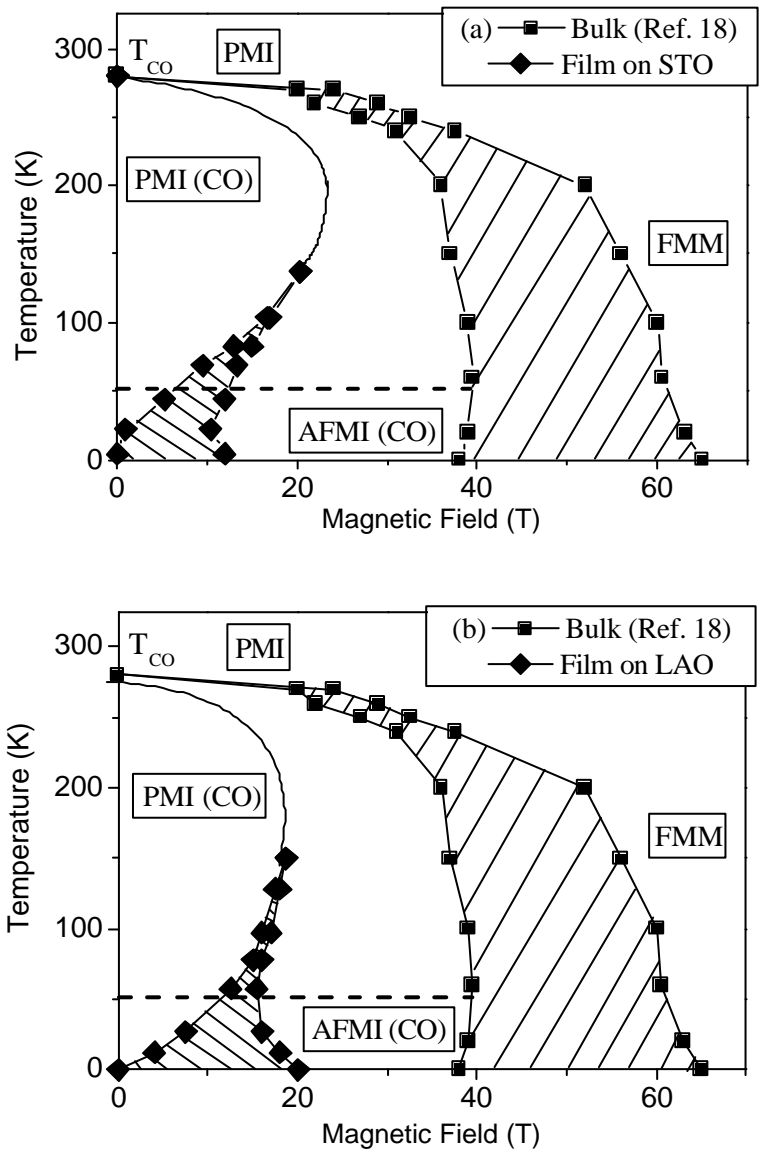

Figure 7

RAUWEL et al.

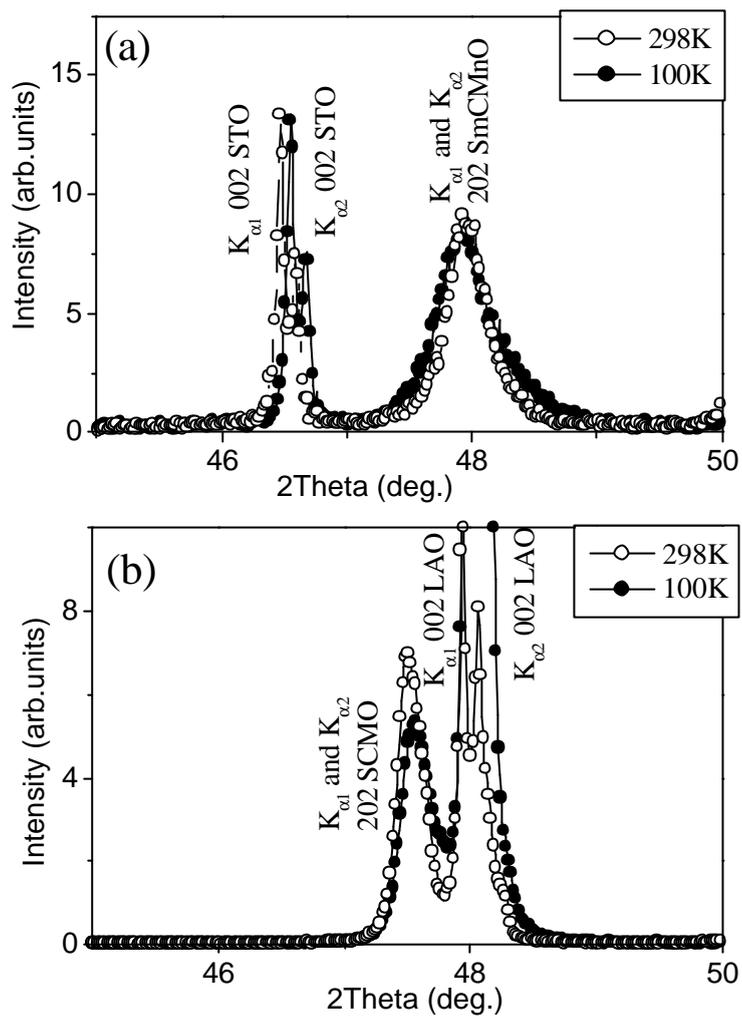

Figure 8

RAUWEL et al.

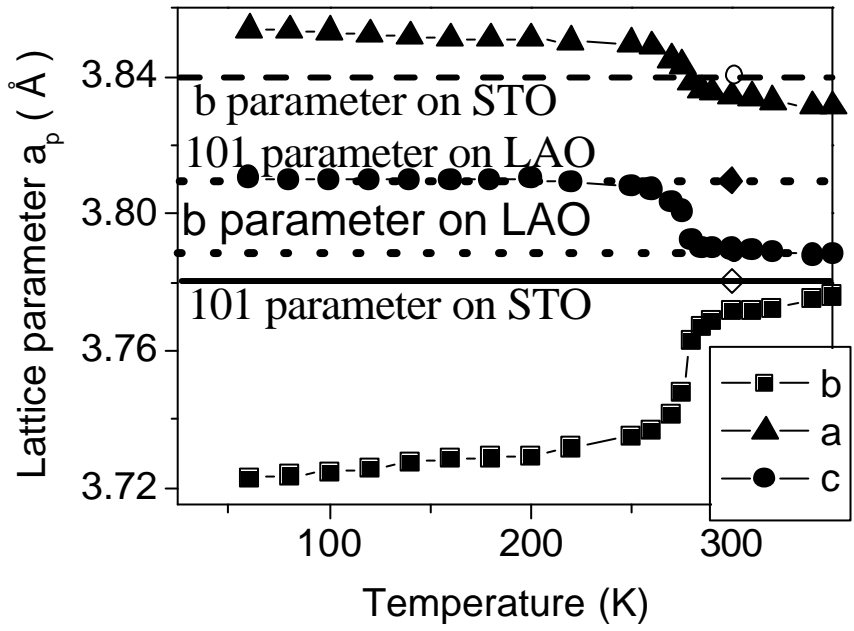

Figure 9

RAUWEL et al.